\documentclass{article}
\usepackage{spconf,amsmath,graphicx}
\usepackage{diagbox}
\usepackage{float}
\usepackage{balance}
\usepackage[T1]{fontenc}
\usepackage[utf8]{inputenc}

\title{Classification of auditory stimuli from EEG signals with a regulated recurrent neural network reservoir}
\name{Marc-Antoine Moinnereau$^{1,2}$, Thomas Brienne$^{1}$, Simon Brodeur$^{1}$, Jean Rouat$^{1}$, Kevin Whittingstall$^{2}$}
\secondlinename{and Eric Plourde$^{1,\star}$\thanks{This project has received funding from NSERC as well as the ACELP-3IT fund of the Université de Sherbrooke.}}
\address{$^{1}$ NECOTIS, Electrical and Computer Engineering Department, Université de Sherbrooke, QC, Canada\\
    $^{2}$ SNAIL, Centre de Recherche du CHUS, Université de Sherbrooke, Sherbrooke, QC, Canada\\
    $^{\star}$ Correspondence : \tt eric.plourde@usherbrooke.ca}
\begin{document}
%
\maketitle
\begin{abstract}
The use of electroencephalogram (EEG) as the main input signal in brain-machine interfaces has been widely proposed due to the non-invasive nature of the EEG.  Here we are specifically interested in interfaces that extract information from the auditory system and more specifically in the task of classifying heard speech from EEGs.  To do so, we propose to limit the preprocessing of the EEGs and use machine learning approaches to automatically extract their meaningful characteristics.  More specifically, we use a regulated recurrent neural network (RNN) reservoir, which has been shown to outperform classic machine learning approaches when applied to several different bio-signals, and we compare it with a deep neural network approach.  Moreover, we also investigate the classification performance as a function of the number of EEG electrodes.  A set of 8 subjects were presented randomly with 3 different auditory stimuli (English vowels a, i and u).  We obtained an excellent classification rate of 83.2\% with the RNN when considering all 64 electrodes.  A rate of 81.7\% was achieved with only 10 electrodes.
\end{abstract}
\section{Introduction}
\label{sec:intro}

The idea of using neural signals to communicate with locked-in persons, who often can't communicate other than through eye movements, has been widely explored \cite{brumberg2010,pei2011,higashi2011}.  In one such setting, a person imagines speaking while his neural activity is recorded.  This activity is then decoded to decipher the intended speech.
Since electroencephalography (EEG) can easily be obtained by simply placing electrodes on the surface of the scalp, its use has therefore been proposed in recent years to classify imagined speech given a fixed set of phonemes or words.  However, the recognition rates of these imagined phonemes or words given EEG measurements have been quite low \cite{higashi2011,dasalla2009,brigham2010}. 
In fact, EEGs are extremely noisy and the neural activity obtained through EEG is not as specific as when using invasive neural measurements since the source of the electrical activity, the neuron, is located relatively far from the recording site.  In fact, better results for the imagined speech task have been achieved with electrocorticography, an invasive measurement approach where the electrodes are directly placed on the surface of the cortex \cite{Martin2016}.

Given the low recognition rates of the imagined speech task with EEGs, we wish here to explore an alternative but related idea by characterizing the task of hearing instead of speaking.  In fact, it has recently been proposed that similar structures in the auditory cortex would be activated when either imagining or hearing a speech and furthermore that the signal representation would also be similar for both cases \cite{Martin2017}.  
  In this study, we will therefore investigate the use of EEG to recognize speech that is heard. This will serve as the gold standard for the much harder task of ''imagining hearing'' a sound, to be investigated subsequently. 

\begin{figure*}[!t]
\includegraphics[scale=0.65]{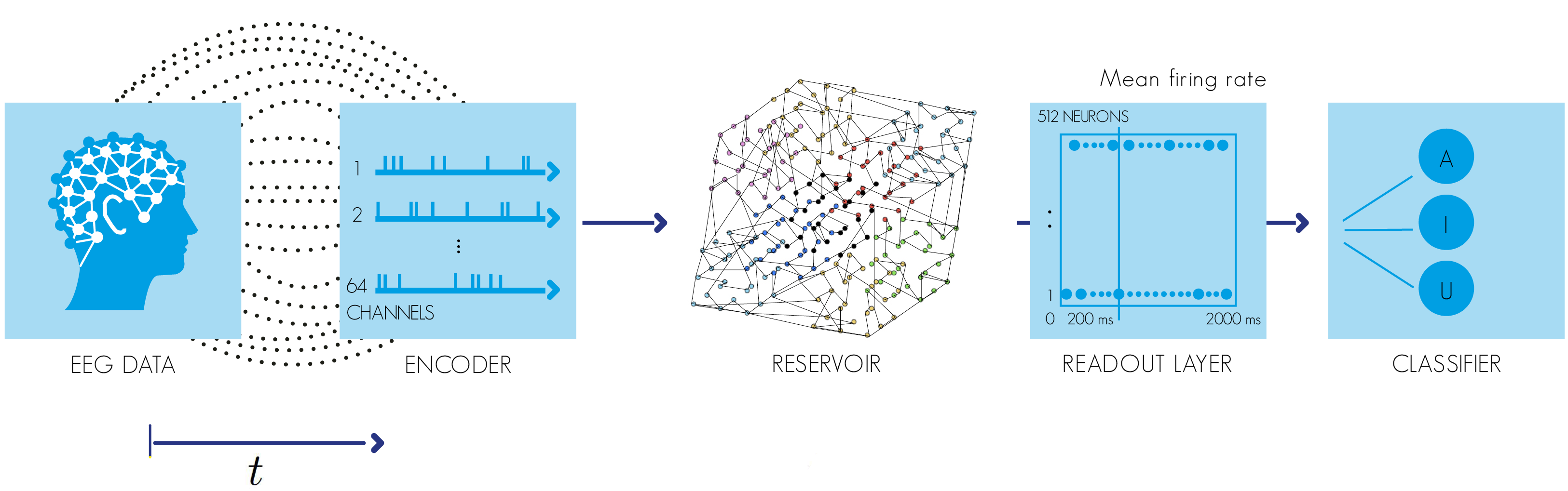}
\caption{Framework of EEG data classification using a RNN reservoir}
\label{architecture}
\end{figure*}

While a great number of studies have reported classification results for imagined speech using EEGs, according to our knowledge, there are surprisingly few studies that have been conducted in order to evaluate the recognition rates that can be achieved when trying to classify heard phonemes or words from EEGs (e.g. \cite{suppes1997,wang2012,Fong2014,Kim2014}).  In \cite{suppes1997}, the EEG for each electrode is optimally band-pass filtered for a given subject in order to maximize the recognition rates.  Even with such an individually optimized approach, the classification rates for seven words, obtained with a support vector machine (SVM) classifier, ranged between 35\% and 53\%.  In \cite{wang2012}, the authors convert the original EEG signal into its surface Laplacian and 2D scalp tangential electric field (LEF).  Also using a SVM, they report a 37.5\% recognition rate for a task involving 8 consonants when using only the temporal signal and a rate of 64.6\% when using the phase of the LEF.   In \cite{Fong2014}, the authors filtered the EEG signals to keep only the delta and theta waves (i.e. between 0.5 Hz and 8 Hz) and report a maximum average classification rate of 77\% between two synthetic vowels.  Another study considered the harder task of classifying perceived speech using only a single trial \cite{Kim2014}.  They used a multivariate empirical mode decomposition prior to performing the classification and obtained rates varying between 68\% and 72\%, depending on the subject, for a classification between two vowels.  In all of these studies, the EEG signal has therefore been significantly preprocessed to extract meaningful parameters associated with the heard speech.     

In this paper, we propose to limit the preprocessing of the EEG to standard band-pass filtering between 0.1 and 45 Hz and use modern machine learning approaches to implicitly extract the important features. In fact, the representation of speech in the cortex is not completely understood and therefore, it is relevant to let the machine learning approaches found the appropriate parameters accounting for this representation.  Moreover, machine learning approaches are excellent at eliminating redundancy in the data which should facilitate data classification.  Many different modern machine learning approaches could have been used to perform this study.  We propose to use a recurrent neural network (RNN) reservoir approach.   In fact, it has been shown that a RNN reservoir gives better classification rates than many machine learning approaches such as the naive Bayes classifier, the multilayer perceptron or the SVM \cite{kasabov}.  The reservoir can approximate any dynamical system and has been used for the classification of electroencephalography (EEG), functional magnetic resonance imaging (fMRI) and magnetoencephalography (MEG) data \cite{kasabov}. 

In this work we use an enhanced and regulated RNN reservoir to classify three vowels from EEGs.  We compare the RNN results with a deep neural network (DNN) approach.  Moreover, we also investigate the classification performance as a function of the number of EEG electrodes used.  We show that the RNN greatly outperforms the DNN and obtains excellent classification results with an average classification rate of 83.2\%.  Furthermore, we show that using only 10 electrodes is sufficient to achieve classification rates that are similar to those obtained with the total 64 electrodes. 


\section{Methodology}
\label{sec:format}

\subsection{Materials and data acquisition}
An Ag/AgCl assembly of 64 surface electrodes was fixed on the actiCAP active electrode system Version I and II (Brain Products GmbH, Germany) and connected to the BrainAmp MR amplifiers. The electrodes were positioned on the scalp according to the 10-20 international standard system \cite{citeulike:9712821}. The pre-processing and encoding in spike sequences of the EEG data were implemented in MATLAB using the EEGlab analysis Toolbox \cite{EEGLAB}; the DNN and RNN were implemented in Python.\\

The experiments were performed on a set of 8 subjects. Conductive gel was placed under each electrode of the cap to enable a suitable contact with the scalp. During the experiments, the subjects kept their eyes closed, avoided facial movements and concentrated on listening to the input audio stimuli. Three English vowels a, i and u (respectively /ej/, /aj/ and /ju/) were chosen as auditory stimuli. An auditory stimulus was presented randomly every 10 s and 1 min breaks were imposed every 5 min so that the subject could relax and stretch. Each stimulus was presented a total of 200 times each. 

\subsection{Data pre-processing}
\label{pre-processing}
The EEG signals were re-sampled to 500 Hz after having been acquired and bandpass filtered between 0.1 and 45 Hz. They were then cut into 2 s intervals where the stimulus had been presented at 0.5 s in each of these intervals. If the amplitude of the signal for a given electrode exceeded the given limit of $\pm$ 75 $\mu$V, the corresponding trials were marked for rejection. Initially, the EEG data was acquired with respect to a common reference site (usually the earlobes or nose) but these references introduced noise. Therefore, the EEG data was re-referenced by a local average \cite{EEGLAB}. Converting data to an average reference is particularly important when the number of electrodes is dense enough to cover the whole head. In fact, the advantage of an average reference is that the sum of the outward positive and negative currents across the entire head will be zero.

\subsection{Regulated RNN reservoir classification system}
The framework of the classification system is presented in Figure \ref{architecture}.  We present the different building blocks of this approach below. 

\subsubsection{Encoder}
In order to be compatible with the RNN reservoir inputs, the EEG signals are transformed into spike trains. Specifically, this encoding is performed using \textit{Ben's Spike Algorithm} (BSA) since it has been found to be highly suitable for EEG data \cite{DBLP}. This algorithm assumes that an analog signal is the result of filtering spikes trains with a reconstruction FIR filter, $h(k)$, as defined in  \cite{schrauwen}.  BSA uses the reconstruction filter impulse response to find the spikes according to the following equations, at every given time $\tau$ : 
\begin{equation} 
e_1(\tau) = \sum_{k=0}^{M} |s(k+\tau)-h(k)|
\label{moneq1}
\end{equation}
\begin{equation} 
e_2(\tau) = \sum_{k=0}^{M} |s(k+\tau)|
\label{moneq2}
\end{equation}
where $s$ is the different EEG signals. If $e_1(\tau) < e_2(\tau) - \theta$, the neuron fires and a spiking value of 1 is associated to that time step, otherwise the spiking activity is represented by a value of 0. The threshold $\theta$ was fixed to $0.9950$ and the FIR filter order was chosen as $M=10$. 

\subsubsection{Regulated RNN reservoir}
A conventional RNN reservoir is composed of a reservoir and a readout layer used for the classification. The reservoir provides computational pre-processing for the subsequent linear readouts. Each neuron in the reservoir receives continuous excitation from the output of the encoder and from the other neurons in the network. The neurons are randomly connected to each other. The reservoir thus includes information about the past and present inputs at a given time. The function of the reservoir is to accumulate temporal and spatial information of a set of input signals into a unique intermediate state with a large dimension to enhance the separation capacity between network inputs \cite{verstraeten2007}. The output of the reservoir is then classified with one layer of neurons for which the connectivity has been found through conventional state of the art supervised learning. \\

We used an unsupervised regulated spiking reservoir that has been specifically designed to adaptively tune the neural network to be on the edge-of-chaos \cite{simon}.  Self-organized criticality is obtained with a local and unsupervised learning rule that has been shown to effectively maintain a long sustained activity while still quickly adapting to the input signal and enhance transients. These characteristics allow for an increase in the separability of the EEG time series \textit{i}) based on their history - events going through time - and \textit{ii}) produced by interdependent processes.  In the reservoir, we used the Leaky Integrate-and-Fire (LIF) neuron model with an adaptive threshold.  Excitatory and inhibitory connections are regulated within the respective ranges of $[0.25, 0.50]$ and $[-0.50, -0.25]$.  Only the readout weights associated to the classifier are trained in a supervised way. The reservoir comprises 512 neurons placed in a three-dimensional grid where 80\% are excitatory and 20\% are inhibitory neurons. The inter-connectivity chosen in the reservoir is a small-world topology as described in \cite{simon}.

\begin{figure*}[t]
\centering
  \begin{tabular}{@{}ccc@{}}
		\includegraphics[scale=0.5]{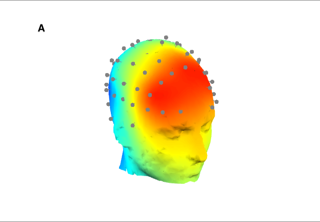} &
		\includegraphics[scale=0.5]{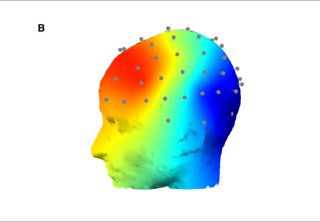} &
		\includegraphics[scale=0.5]{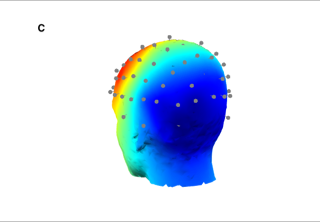} \\
\end{tabular}
\caption{(A) Frontal, (B) lateral and (C) occipital 3D representation of ERPs (average of all stimuli).}
\label{ERP}
\end{figure*}

\subsubsection{Readout layer}
The reservoir accumulates temporal information of all input spike trains and transforms it into dynamic states that can be classified over time. The state of the reservoir is measured with a memoryless readout function that has for input the mean firing rate of each neuron in the reservoir. The mean firing rates are estimated on 200 ms duration frames. 

\subsubsection{Classifier}
Finally, the classification of the output of the readout layer into class labels associated to each input vowel (e.g. 0 for a, 1 for i and 2 for u) is done through simple linear regression.

\subsection{DNN}
The DNN used for comparison is a standard convolutional network (CNN) with maximum pooling non-linearity after each convolutional layer. It is composed of 3 convolutional layers of 64 filters with rectified-linear activations (ReLU), a temporal width of 3 and a stride of 1. The input to the neural network is the 64 pre-processed EEG signals coming from the electrodes. The last convolutional feature map is connected to a dense layer of 32 hidden neurons with ReLU activations. A final dense layer of 3 output neurons with softmax activations provides an output for each class. The metaparameters of the networks (e.g. number of layers and hidden neurons) were optimized using grid search in order to maximize the classification rates obtained. The neural network was trained using the Adam back-propagation algorithm \cite{Kingma2015} to minimize a categorical cross-entropy loss function for 100 epochs. The same training dataset and methods (e.g. cross-validation) as the reservoir was used.

\begin{table}[t]
\caption{Combined classification rates of a, u and i with 8 subjects for both the DNN and RNN using all 64 channels.} 
\centering
\begin{tabular}{c|c|c}
Subjects & DNN & RNN  \\ \hline
1 & 55.7\% & 88.8\% \\ 
2 & 42.9\% & 67.7\%  \\
3 & 45.6\% & 87.7\%  \\
4 & 58.4\% & 84.6\%  \\
5 & 70.8\% & 78.5\%  \\
6 & 43.7\% & 79.3\%  \\
7 & 30.8\% & 88.2\%  \\
8 & 40.6\% & 90.4\%  \\ \hline \hline
Average & 48.6\%  & 83.2\% \\
\hline
\end{tabular}
\label{resultDNNRNN}
\end{table}

\begin{table}[t]
\caption{Classification rates of a, i and u for the RNN using all 64 channels (mean of 8 subjects).} 
\centering
\begin{tabular}{c|c|c|c|c} \cline{1-5}
Vowel & a & i & u & all \\ \hline
Classification rate & 81.1\% & 87.0\% & 81.3\% & 83.2\% \\
\hline
\end{tabular}
\label{resultvowels}
\end{table}


\begin{table}[t]
\centering
\caption{Combined classification rates of a, i and u as a function of the number of electrodes for the RNN (mean of 8 subjects).} 
\begin{tabular}{c|c|c|c|c} \cline{1-5}
No. of electrodes & 1 & 3 & 10 & all \\ \hline
Classification rate & 57.3\% & 71.4\%  &	81.7\% & 83.2\% \\
\hline
\end{tabular}
\label{resultchannel}
\end{table}


\section{Results and discussion}

To validate the EEG signal acquisition during the auditory task, we first computed the event-related potential (ERP) presented in Fig. \ref{ERP}.  The observed neural activity is as expected for an auditory task since we can easily observe the characteristic strong activity in the frontal and occipital cortical regions.

The classification results obtained for both the DNN and the regulated RNN approaches are presented in Table \ref{resultDNNRNN}.  We used a 5-fold cross-validation for all the results presented.  The pre-processing step described earlier typically rejected 30\% of the epochs.  A vowel was accurately classified if the class label yielded by the classifier corresponded to the input auditory stimuli having been presented.  All 64 electrodes were used.  As can be observed, the classification rates obtained with the RNN highly exceeds those of the DNN.    The best averaged rate of classification for the RNN is obtained for subject 8 with 90.4\% while the worst is reported for subject 2 with 67.7\% denoting a certain variability through subjects, which is expected in EEG measurements.  Table \ref{resultvowels} presents the classification results for each of the vowels independently for the RNN approach.  It is interested to note the robustness of the approach where all classification rates are well above the chance level of 33.3\% with a slightly better result for the vowel i.  It’s hard to compare these results with the previous work where many different experimental conditions (e.g. different type and number of stimuli) and preprocessing has been used.  However, we show here that excellent classification results can be obtained with minimal preprocessing of the EEGs.

Moreover, the recognition rates should improve by increasing the number of electrodes used in the classification.  In fact, the neural signal originating from the auditory cortex at a specific location can be thought to spread to the scalp and be distributed among different EEG channels.  Table \ref{resultchannel} shows the best results when using the $N$ best electrodes to perform the classification.  We observe that, indeed, the classification results improve with an increasing number of electrodes.  However, with only 10 electrodes the classification rate is already quite close to the one achieved with all 64 electrodes. 


\section{Conclusion}
In this paper, we have proposed to use a machine learning approach (i.e. a regulated RNN reservoir) to automatically extract the meaningful characteristics of minimally preprocessed EEGs and perform the classification of three vowels.  Even with limited preprocessing, we obtained an excellent average classification rate of 83.2\%.  Moreover, a small number of electrodes is necessary to achieve high classification rates where only a 1.5\% difference in classification rates are observed when using only 10 instead of all 64 electrodes.     

\balance

\bibliographystyle{IEEEbib}
\bibliography{strings}

\end{document}